\documentclass[twocolumn,showpacs,prl,aps]{revtex4}
\usepackage{epsfig}
\def\dt{\Delta t}
\def\dx{\Delta x}

\def\grad{\nabla}
\def\lk{\lambda_{\bf k}}

\begin{document}
\title{Maximally-fast coarsening algorithms}
\author{Mowei Cheng}
  \email{mowei@fizz.phys.dal.ca}
\author{Andrew D. Rutenberg}
  \homepage{http://www.physics.dal.ca/~adr}
  \affiliation{Department of Physics and Atmospheric Science, Dalhousie University, 
Halifax, Nova Scotia, Canada B3H 3J5}
\date{\today}

%%%%%%%%%%%%%%%%%%%%%%%%%%%%%%%%%%%%%%%%%%%%%%%%%%%%%%%%%%%%%%%%%%
\begin{abstract}
We present maximally-fast numerical algorithms for conserved coarsening systems that are 
stable and accurate with a growing natural time-step $\Delta t=At_s^{2/3}$. For non-conserved 
systems, only effectively finite
timesteps are accessible for similar unconditionally stable algorithms. We compare the scaling
structure obtained from our maximally-fast conserved systems directly against
the standard fixed-timestep Euler algorithm, and find that the error scales as 
$\sqrt{A}$ --- so arbitrary accuracy can be achieved. 
\end{abstract}


%%%%%%%%%%%%%%%%%%%%%%%%%%%%%%%%%%%%%%%%%%%%%%%%%%%%%%%%%%%%%%%%%%
\maketitle

Phase-ordering kinetics studies the evolution of structure after a quench from a 
disordered phase into an ordered phase.  
The later stages of most phase-ordering processes in simple systems show universal scaling behavior 
described by a single growing length scale which increases as a power law in time, 
$L(t) \sim t^\alpha$, where $0 < \alpha \leq 1$ \cite{Brayreview}.  
For the scalar order-parameter systems considered
in this paper, $\alpha=1/2$ and $1/3$ for non-conserved and conserved dynamics, respectively 
\cite{Brayreview}.  While these growth exponents and their universality can be understood 
in terms of interfacial motion leading to domain coarsening \cite{energyscaling}, the 
time-independent scaled structure that result is less well understood. 

Computer simulation is an effective technique to systematically study these 
non-linear non-equilibrium coarsening systems.  To maintain accuracy, the 
discretized dynamics must move interfaces {\em at most}
a small fraction of the interfacial width, $\xi$, in a single 
timestep $\Delta t$. This determines a maximal or natural timestep of coarsening systems, 
$\dt_{nat} \sim \xi/ (dL/dt) \sim t^{1-\alpha}$, that
grows in time.  Unfortunately common time-discretizations are 
unstable for timesteps above a fixed threshold determined by the lattice spacing $\dx$ 
\cite{checkerboard}.  Any such fixed timestep algorithm is
increasingly inefficient at late times compared to the natural timestep.  
Various algorithms have been proposed to 
make simulations more efficient, including the cell-dynamical-scheme \cite{CDS} and Fourier 
spectral methods \cite{fourierspectral}. However, these 
approaches still require a fixed time-step for numerical stability.  

There is a newly developed class of unconditionally stable semi-implicit algorithms 
\cite{Eyre,VR} that impose no stability constraints on the timestep $\dt$, which is 
then determined by accuracy considerations.  Since we generally expect larger $\dt$ to 
lead to larger errors, there is a tradeoff between speed and accuracy. This tradeoff is 
best resolved by picking growth rates for $\dt$ that 
induce an error in the correlations that is approximately constant in magnitude throughout 
the scaling regime, where the magnitude can be chosen to be less than other systematic
sources of error such as initial transients or finite-size effects. 
While errors of {\em single} growing timesteps
can be small \cite{VR}, this begs the question of how much those single-step errors accumulate 
in correlations at late times after the quench. For example, we might expect that some 
types of single-step errors would be benign, given the irrelevance of small 
amounts of random thermal noise to the scaled structure \cite{Brayreview}.  
Nevertheless, we would expect timesteps growing faster than $\dt_{nat}$ to lead to 
unacceptable levels of error.  

In this paper, we compare the scaled correlations of unconditionally stable ``Eyre'' dynamics 
driven with a growing timestep $\dt$ with the correlations evolved
with an explicit Euler update. The latter, while slow, provides an arbitrarily accurate reference 
at late times. For conserved Cahn-Hilliard dynamics \cite{CH}, we find that the maximal 
difference of the scaled structure has a constant magnitude in the scaling regime when the 
Eyre algorithms are driven at the natural or maximal timestep 
\begin{equation}
	\dt=At_s^{2/3}, 
	\label{eq:natural}
\end{equation} 
for conserved dynamics. Here $t_s$ is a natural ``structural time'' determined by the
decreasing system energy (see Eq.~(\ref{eq:dtsdefine}) below).
We find that this correlation error scales as $\sqrt{A}$ for small $A$, and so can be made 
arbitrary small while retaining a {\em maximally fast} conserved coarsening algorithm which
corresponds to moving interfaces a finite fraction of the interfacial width in every timestep.
With a similar class of unconditionally stable algorithms for 
non-conserved dynamics \cite{VR}, we find that only a fixed factor speedup 
is possible compared to the Euler algorithm as measured by the structural time $t_s$.
With a Fourier-space analysis of the dynamics, we explain how 
the natural timestep can accurately be used in conserved coarsening while 
only a fixed timestep is possible for non-conserved coarsening.  

Cahn-Hilliard dynamics of a conserved scalar field $\phi({\bf r},t)$ are 
\begin{equation}
\partial \phi/ \partial t =\grad^2 \delta F / \delta \phi=-\grad^2(\phi+\grad^2\phi-\phi^3), 
	 \label{eq:CHbasic}
\end{equation}
where $F \equiv \int d^dx [(\grad \phi)^2 + (\phi^2-1)^2/4]$ is the free energy 
in $d$ spatial dimensions with a double-well potential corresponding to two distinct 
ordered phases at $\phi = \pm 1$.  
These dynamics can be semi-implicitly discretized in time by 
\begin{eqnarray}
	\tilde{\phi}_{t+\dt} + (1-a_1) \dt \grad^2 \tilde{\phi}_{t+\dt} + (1-a_2) \dt 
	\grad^4 \tilde{\phi}_{t+\dt} \nonumber \\
	= \phi_t - \dt \grad^2 (a_1 \phi_t + a_2 \grad^2\phi_t - \phi_t^3),
	\label{eq:CHdirect}
\end{eqnarray}
where the discretized dynamics are unconditionally stable for any $\dt >0$ when
 $a_1>2$ and $a_2<0.5$ \cite{VR}. 
This equation implicitly defines the updated field, $\tilde{\phi}_{t+\dt}$, and can be 
directly solved in Fourier space to give an Euler-like update
\begin{equation}
\tilde\phi_k(t+\dt)= \phi_k(t) + \dt_{eff}(k) \dot{\phi}_k
	\label{eq:CHdteff}
\end{equation}
where the $k$-dependent effective timestep is
\begin{equation}
	\dt_{eff}(k) 	\equiv \dt/(1-\dt K), 
	\label{eq:dteff}  
\end{equation}
where $\lk$ is the Fourier-transformed Laplacian ($\lk=-k^2$ in
the continuum limit),  and
\begin{equation}
	K \equiv (a_1-1) \lk + (a_2-1) \lk^2. 
\end{equation}
We note that $K <0$ for unconditionally stable algorithms.

Our numerical work is done in two-dimensions ($2d$), with systems of linear size
$L_\infty =256$ (at least $200$ samples) and $L_\infty =512$ (at least $20$ samples).  
We use a lattice spacing $\dx=1$ and periodic boundary conditions.
For Euler discretizations of the conserved dynamics Eq.~(\ref{eq:CHdirect}) with $a_1=a_2=1$, 
we use $\dt=0.03$, while for unconditionally stable discretizations we use $a_1=3$ and $a_2=0$. 

The $k$-dependent effective timestep in Eq.~(\ref{eq:dteff}) led us to investigate the actual
timestep taken for a given algorithmic timestep $\dt$. To do this we exploit the 
power-law decay of the free-energy density, $\epsilon \equiv F/V$, in the late-time scaling 
regime, $\epsilon \sim 1/L \sim t^{-\alpha}$ \cite{energyscaling}, to introduce the ``structural'' time 
\begin{equation}
	t_s \equiv B \epsilon^{-1/\alpha},
	\label{eq:dtsdefine}
\end{equation}
where $\alpha =1/3$ or $1/2$ for conserved or non-conserved coarsening, respectively, and 
$B$ is chosen so that $\dt_s=\dt$ for small $\dt$ in the late-time scaling regime. For
conserved dynamics we use $B=0.286$, while for non-conserved $B_{nc}=0.105$. 
The evolution of the structural time allows us to measure the real speed-up of our 
coarsening algorithms, and it will also provide important insight into errors of the scaled 
structure.  In general we find $\dt_s/\dt \leq 1$ (see Fig.~\ref{FIG:error} and 
\ref{FIG:nonconserved} below), and so we conservatively use the structural time $t_s$ to drive 
our algorithms, as in Eq.~(\ref{eq:natural}). 

\begin{figure}[htb]
\begin{center}
\includegraphics[width=260pt]{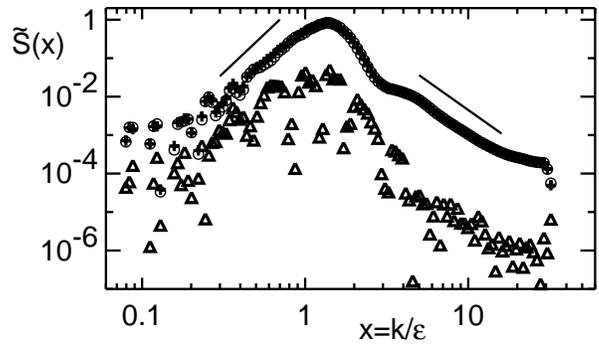}
\caption{The average scaled structure $\tilde{S}(x) \equiv \epsilon^2 S(x \epsilon)$ vs. 
$x \equiv k/\epsilon$ for $L_\infty =512$ system at $t_s=1024$ with the Euler update (circles) 
with $\dt = 0.03$ and the Eyre update (``$+$'') with $\dt= A t_s^{2/3}$ where $A=0.01$.
Triangles indicate the absolute difference between Eyre and Euler updates, $\Delta \tilde{S}(x)$.}
\label{FIG:structure}
\end{center}
\end{figure}

We find that using the natural timestep, Eq.~(\ref{eq:natural}), leads to accurate correlations
in the scaling regime --- as compared to systems evolved with a simple Euler time-discretization.
We measure $S(k,t)= \langle \phi_k \phi_{-k} \rangle$, where the angle-brackets indicate an average
over orientations and initial conditions.  We obtain the scaling form using the energy density, 
$\epsilon$, so that $\tilde{S}(x) \equiv \epsilon^2 S(x \epsilon)$ is a scaling function of 
$x \equiv k / \epsilon$.  In Fig.~\ref{FIG:structure} we plot $\tilde{S}(x)$ vs. $x$ to  
illustrate the excellent overlap between Euler (circles) and Eyre
(``$+$'') dynamics with $\dt=A t_s^{2/3}$ and $A=0.01$.  To quantify the error we 
take the absolute value of the maximal value of the difference between the
scaled structures (shown for general $k/\epsilon$ with triangles).   We find that this
maximum difference is approximately constant in magnitude throughout the scaling regime. 

We average this absolute error over the scaling regime, as determined by the scaling collapse of 
the scaled structure.  We observe a small $A$-dependent difference between the Euler algorithm and 
the naturally driven Eyre algorithm, as shown in Fig.~\ref{FIG:error}.  By repeating the measurement for 
different system sizes, both $L_\infty =256$ (open circles, averaged over 4 times in $t_s \in [60,190]$) 
and $L_\infty =512$ (open squares, averaged over 9 times in $t_s \in [60,1500]$), we confirm that
finite-size effects are not significant.  We have enough independent Euler samples that no baseline
errors due to residual stochastic effects are seen --- the errors shown are the systematic error due to 
$A$.  We observe an approximately $\sqrt{A}$ dependence on the average error.   
This implies that arbitrarily accurate measurements of the scaled structure can be made with 
maximally driven Eyre algorithms. 

\begin{figure}[htb]
\begin{center}
\includegraphics[width=260pt]{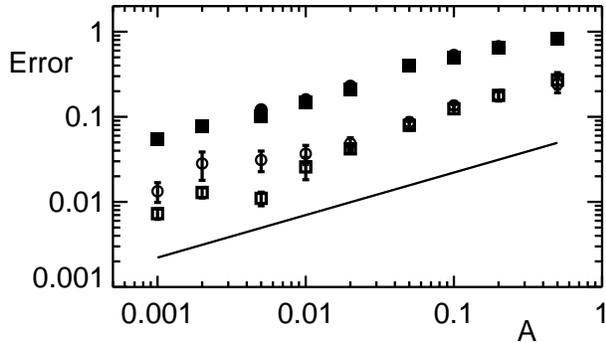}
\caption{The Eyre structural error vs. $A$ for Eyre algorithms driven at the natural
timestep $\dt=A t_s^{2/3}$. The error is the maximum absolute difference between
the Eyre correlations and the Euler correlations. Shown is data for $L_\infty =256$ (open
squares) and $L_\infty =512$ (open circles). Also shown with filled symbols are the asymptotic
values of $1-\dt_s/\dt$ for various $A$. Both errors exhibit $\sqrt{A}$ behavior, as 
discussed in the text.} 
\label{FIG:error}
\end{center}
\end{figure}

From the effective timestep, Eq.~(\ref{eq:dteff}), 
we can see qualitatively why the natural timestep, Eq.~(\ref{eq:natural}), 
is accurate for conserved updates. Since Fourier modes  
$k \lesssim 1/L \sim t^{-1/3}$ correspond to the domain structure, then for the natural 
timestep with very small $A$ the effective timestep is only $k$-dependent deep in the Porod 
tail where $kL \gg 1$.  However, the Porod tail \cite{Brayreview} is simply the reflection of the 
amount of interface in the system at $k L \approx 1$. 
Systematic errors will arise as $A$ gets larger and the 
$k$-dependence of $\dt_{eff}$ becomes more significant at $kL \approx 1$.  

Previous real-space single-step error analysis \cite{VR} indicated that 
the natural timestep would not be accurate for direct Eyre algorithms like
Eq.~(\ref{eq:CHdirect}) because of growing errors close to interfaces \cite{correctVR}.  We have
examined interfacial {\em profiles} of naturally driven algorithms, 
as probed by the $k \sim \xi^{-1}$ tail of the unscaled structure, and
found quantitative agreement with Euler algorithm (see large $x$ data in Fig.~\ref{FIG:structure}). 
Single-step errors must then correspond to errors in the interfacial {\em motion}, which 
can be probed by considering the difference between the algorithmic timestep $\dt$ and the 
resulting structural timestep $\dt_s$.  We find that $\dt_s/\dt<1$ in general, which indicates 
reduced interfacial motion.  In Fig.~\ref{FIG:error} we show with filled symbols 
that $1-\dt_s/\dt$ scales as $\sqrt{A}$ for small $A$.  We recover this asymptotic result 
using an energy-scaling argument \cite{energyscaling,Brayreview} in general spatial dimension $d$.

From Eq.~(\ref{eq:dtsdefine}), with $\alpha=1/3$ for conserved dynamics, 
we have $\epsilon = B^{1/3} t_s^{-1/3}$ so that 
\begin{equation}
\dt_s = -3B^{-1/3} \Delta \epsilon t_s^{4/3}.
	\label{eq:CHdts}
\end{equation}
On the other hand we can integrate the energy dissipated in one timestep for each 
Fourier component \cite{energyscaling},
\begin{eqnarray}
\Delta \epsilon &\simeq& \int d^dk /(2\pi)^d
			\langle (\delta F/\delta \phi_k) \Delta \phi_k \rangle, \\
		&=&-\int d^dk/ (2\pi)^d k^{-2} \dt_{eff}(k,\dt) T_k
	\label{eq:CHde}
\end{eqnarray}
where the second line uses the time-derivative $\dot{\phi}_{-k}= -k^2 \delta F/\delta \phi_k$ from
Eq.~(\ref{eq:CHbasic}), $\Delta \phi_k = \dt_{eff} \dot{\phi}_k$ from Eq.~(\ref{eq:CHdteff}),
and the time-derivative scaling function
\begin{eqnarray}
	T_k \equiv \langle \dot{\phi}_k \dot{\phi}_{-k} \rangle 
			= \dot{L}^2 L^{d-2} h(kL) 
	\label{eq:tk}
\end{eqnarray}
where the scaling form is shown.  $T_k$ is expected to have a 
Porod-like $h(x) \sim x^{1-d}$ tail for $x \gg 1$ \cite{energyscaling} and this 
has been observed in $d=2$ \cite{tobepublished}.  With these asymptotics and 
$\dt_{eff}$ from Eq.~(\ref{eq:dteff}), the $\Delta \epsilon$ integral
converges and hence becomes time-independent as the UV cutoff $O(L/\xi)$ becomes large. 
For $\dt=At_s^{2/3}$ we obtain 
\begin{eqnarray}
\dt_s/\dt \propto \int_{0}^\infty dx x^{d-3} h(x)/(1+A'x^2) 
	\label{eq:CHdtdts}
\end{eqnarray}
where $A' \equiv A(a_1-1)/L_0^2$ and $L=L_0 t_s^{1/3}$ in the scaling regime.  For small $A$, the
leading contribution is $O(\sqrt{A'})$ from the large $x$ regime.  Since $\dt_s/\dt=1$ in
the limit of $\dt \rightarrow 0$ when $A = 0^+$, we have 
\begin{eqnarray}
	1-\dt_s/\dt \propto \sqrt{A'} + O(A')
	\label{eq:dtsdt}
\end{eqnarray} 
This behavior is observed, as shown by the filled points in Fig.~\ref{FIG:error}.

What is the connection, if any, between the timestep error $1-\dt_s/\dt$ and the structural 
error, both of which exhibit $\sqrt{A}$ dependence at small $A$ for natural timesteps?  
As discussed before, $1- \dt_s/\dt \sim \sqrt{A}>0$ indicates an error of (reduced) interfacial
motion. It is reasonable that this error shows up in the correlations at the same order, $O(\sqrt{A})$.
It is interesting that this error accumulates into a constant contribution to the scaled correlations
within the scaling regime. 

We now consider non-conserved Allen-Cahn coarsening dynamics, which are governed by
\begin{equation}
\dot{\phi}=-\delta F / \delta \phi=\phi+\grad^2\phi-\phi^3.
	 \label{eq:ACbasic}
\end{equation}
These dynamics can be semi-implicitly discretized in time by 
\begin{eqnarray}
	\tilde{\phi}_{t+\dt} + (a_1-1) \dt \tilde{\phi}_{t+\dt} + (a_2-1) \dt 
	\grad^2 \tilde{\phi}_{t+\dt} \nonumber \\
	= \phi_t + \dt (a_1 \phi_t + a_2 \grad^2\phi_t - \phi_t^3),
	\label{eq:ACdirect}
\end{eqnarray}
where the discretized dynamics are unconditionally stable for any $\dt >0$ when
 $a_1>2$ and $a_2<0.5$ \cite{VR}. 
In the same spirit as conserved dynamics, we obtain an effective timestep  
\begin{equation}
	\dt_{eff}(k,\dt) =\dt/(1-\dt N),
	\label{eq:ACdteff}
\end{equation}
where $N \equiv (1-a_1) + (1-a_2) \lk$ and $N<0$ for stable algorithms with $a_1>2$ and $a_2<0.5$ 
\cite{VR}.  This directly implies that 
$\dt_{eff} \leq 1/(1-a_1)$ even when $\dt \rightarrow \infty$, so that this class
of unconditionally stable non-conserved algorithms effectively cannot be accelerated. 
We confirm this by calculating and measuring $\dt_s$ when $\dt=\infty$. 

\begin{figure}[htb]
\begin{center}
\includegraphics[width=260pt]{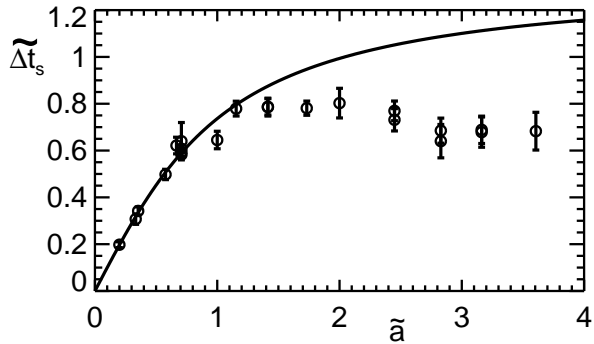}
\caption{For non-conserved update driven with $\dt=\infty$, we plot 
$\widetilde{\dt}_s \equiv \dt_s \sqrt{(a_1-1)
(1-a_2)}$ vs. $\tilde{a} \equiv \sqrt{(1-a_2)/(a_1-1)}$ showing the prediction 
(solid line with $\xi=0.85$)) of Eq.~(\protect\ref{eq:dtsNC2}).  
}
\label{FIG:nonconserved}
\end{center}
\end{figure}

We adapt the development in Eq.~(\ref{eq:CHdts}) - (\ref{eq:CHdtdts}) for non-conserved
dynamics.  We have $\epsilon = B_{nc}^{1/2} t_s^{-1/2}$ so that 
\begin{eqnarray}
\dt_s = -2B_{nc}^{-1/2} \Delta \epsilon t_s^{3/2}.
	\label{eq:ACdts}
\end{eqnarray}
Integrating the energy dissipated in one timestep, 
\begin{equation}
	\Delta \epsilon \simeq -\int d^dk (2\pi)^{-d} \dt_{eff}(k,\dt) T_k
	\label{eq:ACde}
\end{equation}
where we use $\dot{\phi}_{-k} = - \delta F/\delta \phi_k$.
Using Eq.~(\ref{eq:ACdteff}) and Eq.~(\ref{eq:tk}) with $L = L_0 t_s^{1/2}$ and 
$\lk = -k^2$ we can solve for $\dt_s$ when $\dt=\infty$: 
\begin{eqnarray}
	\dt_s & = &  \frac{L_0 L}{ 4 \pi \sqrt{B_{nc}}} 
		\int_{0}^{L/\xi} \frac{x^{d-1} h(x) dx}{(a_1-1)L^2+ x^2(1-a_2)} 
	\label{eq:dtsNC1}\\
		&=& \xi \frac{\tan^{-1}\Big(\sqrt{\frac{1-a_2}{a_1-1}}\frac{1}{\xi}\Big)}
			{\sqrt{(a_1-1)(1-a_2)}},
        \label{eq:dtsNC2}
\end{eqnarray}
where we take the late-time $L \rightarrow \infty$ limit in the second line and have
used $h_{nc}(x) \sim x^{1-d}$ for $x \gg 1$ \cite{energyscaling,Brayreview,tobepublished},
without which $\dt_s$ is not time-independent. The
overall $\xi$ factor on the second line comes from imposing $\dt_s=\dt$ for small $\dt$ to 
determine $B_{nc}$, before the $\dt \rightarrow \infty$ limit is taken. We find a 
constant $\dt_s$ that depends 
only on $a_1$ and $a_2$, as well as the inverse UV cutoff $\xi$.  In Fig.~\ref{FIG:nonconserved}, 
for $d=2$, we plot the measured asymptotic $\widetilde{\dt}_s \equiv \dt_s \sqrt{(a_1-1)(1-a_2)}$ 
(averaged over the time-independent scaling regime with {\em variances} shown) 
vs. $\tilde{a} \equiv \sqrt{(1-a_2)/(a_1-1)}$. We observe data collapse for a variety 
of $a_1$ and $a_2$, as suggested by Eq.~(\ref{eq:dtsNC2}). However
we only agree with the calculated functional form for small $\tilde{a}$ -- the solid
line indicates best fit by eye with $\xi=0.85$.  The discrepancy appears to be due to 
$\tilde{a}$ dependence of the time-derivative correlations (data not shown), 
indicating significant systematic errors when 
$\dt \rightarrow \infty$ despite the finite effective timestep.  

In summary, we have shown that for conserved dynamics we obtain accurate scaled
correlations for maximally fast algorithms with $\dt=At_s^{2/3}$.  
The structural error behaves as $\sqrt{A}$ for small $A$.  These results are consistent
with previous real-space single-step error-analysis \cite{VR} away from interfacial regions.
For these maximally fast algorithms the relative speedup with respect to a fixed timestep is of 
the order $(L_\infty/\dx)^d$, where $L_\infty/\dx$ is the discretized linear system size.   
Maximally fast algorithms provide the most efficient means to reach the scaling limit for 
large systems. 

Surprisingly, a similar class of algorithms does not lead to acceleration of non-conserved 
dynamics. Only a constant time-step is observed, as measured by the structural 
time, even when the unconditionally stable algorithm is driven with $\dt \rightarrow \infty$.  

We expect that these unconditionally-stable semi-implicit algorithms find broader
application. In the systems investigated so far, the regime of unconditional stability coincides
with the easily determined regime of linear-stability. 
For coarsening systems that exhibit a growing natural timestep, large accelerations are
possible with growing $\dt$. While the non-conserved dynamics cautions us that effective
acceleration is not always produced, it also illustrates the diagnostic value of the 
ratio $\dt_s/\dt$. Indeed, $1-\dt_s/\dt$ appears to provide a good proxy for systematic structural 
errors in the scaling regime. 

We thank the Natural Science and Engineering Research Council of
Canada and the Canadian Foundation for Innovation for support.  
We thank Ben Vollmayr-Lee for valuable ongoing discussions. 

%%%%%%%%%%%%%%%%%%%%%%%%%%%%%%%%%%%%%%%%%%%%%%%%%%%%%%%%%%%%%%%%

%\vspace*{-5mm}

\end{document}